# Estimation of the Total Mass of 10 Exoplanets and their Host Stars Based on the Primary Transit Method

Poro A.[1,2], Hedayatjoo M.[1,2], Dashti Y.[1,2], MohammadiZadeh F.[1,2], Hashemi M.[1,2], Rajaei E.[1,2], Kazemi A.[2], Sarostad A.[2], Nastaran M.[2], Zarei Z.[2], Dehghani Ghanatghestani A.[2]

[1]The International Occultation Timing Association-Middle East section, info@iota-me.com
[2]Exoplanet Transit Project (ETP-1), Transit department, IOTA-ME, Iran

**Abstract**

In this study, ten exoplanets were studied. Their photometric observations were obtained from the ETD. After performing the data reduction steps, their parameters were obtained through Exofast online software. Then the total mass of the exoplanets and the host stars were obtained through a software that was mentioned in this study. The location of the host stars in this study was also plotted in the H-R diagram.

*Keywords: Photometry, EXOFAST, Mass, Metallicity*

## Introduction

Ever since the first discoveries were made about exoplanets in 1992, many researchers have been studying their properties by using data provided from different sources (Swift, 2010). A part of these sources is the ground-based observations, conducted in small observatories around the world. Even though space missions such as TESS are able to survey a great percentage of the sky and provide us with more reliable data, these observatories show great potential for producing data on brighter stars and have contributed immensely to this field (Davoudi, 2020).

Throughout this article, we analyzed the parameters of the following exoplanets: CoRoT-12 b, HAT-P-52 b, HAT-P-57 b, HATS-28 b, HATS-34 b, KELT-3 b, WASP-61 b, WASP-67 b, WASP-122 b, and WASP-140 b. All of the planets were discovered through the primary transit method, between the years 2010 and 2016. Their orbital periods range from 1.710 to 4.674 days, and their host stars' apparent magnitudes are between 9.8 to 15.52 (The Extrasolar Planets Encyclopaedia[1]).

In order to calculate their parameters and carry out a detailed analysis, we got the raw data from the Exoplanet Transit Database (ETD[2]). Then we proceeded with data reduction using the Phoebe 0.32 software[3]. Therefore, we gave the normalized data to EXOFAST[4] software and obtained the parameters for the planets. The purpose of this study is to make a comparison of some of the exoplanets' parameters with other studies.

## Data sets

Exoplanet Transit Database (ETD)[5] the project of the Czech Astronomical Society, has started working since September 2008 as a website containing data of light curves of exoplanets and variable stars observed by professional and also amateur astronomers who have to spend a long period of time observing short-term and long-term changes of a transit (Poddaný, 2010). The majority of exoplanets and variable stars that have already been found are gathered in ETD website which is made of three parts, the first one Transit predictions the second one Model fit the data and the third one O-C plots that show the transit predictions, Transit Timing Variation (TTV) variation of depth and the duration (Poddaný, 2010). In addition, before ETD came online, merely two transit databases could be used. The Amateur Exoplanet Archive (AXA)[6] and NASA Star and Exoplanet Database (NStED)[7] that unfortunately were not as comprehensive as ETD. ETD has used these two websites and the TRESCA[8] project as sources to make the database where researches can get data from or add information to (Poddaný, S, 2010). In Table 1 shows specifications of the host star and the planets that studied.

---

[1] http://exoplanet.eu
[2] http://var2.astro.cz/ETD
[3] http://phoebe-project.org/1.0/download
[4] https://exoplanetarchive.ipac.caltech.edu/cgi-bin/ExoFAST/nph-exofast
[5] http://var2.astro.cz/ETD/

**Table 1. Specifications of the host stars and the planets. The exoplanets detected by Primary Transit method.**

| Star Name | RA$_{2000}$ | Dec$_{2000}$ | D. (pc) | Spect. | App. Mag. | Planet | Discovered |
|-----------|-------------|--------------|---------|--------|-----------|--------|------------|
| CoRoT-12  | 06 43 03.76 | -01 17 47.14 | 1150    | G2V    | 15.52     | CoRoT-12 b | 2010   |
| HAT-P-52  | 02 50 53.20 | +29 01 20.52 | 385     | -      | 14.07     | HAT-P-52 b | 2015   |
| HAT-P-57  | 18 18 58.42 | +10 35 50.12 | 303     | -      | 10.47     | HAT-P-57 b | 2015   |
| HATS-28   | 18 57 35.92 | -49 08 18.55 | 521     | G      | -         | HATS-28 b  | 2016   |
| HATS-34   | 00 03 05.87 | -62 28 09.61 | 532     | -      | 13. 85    | HATS-34 b  | 2016   |
| KELT-3    | 09 54 34.38 | +40 23 16.97 | 178     | F      | 9.8       | KELT-3 b   | 2012   |
| WASP-61   | 05 01 11.91 | -26 03 14.96 | 480     | F7     | 12.5      | WASP-61 b  | 2011   |
| WASP-67   | 19 42 58.52 | -19 56 58.52 | 225     | K0V    | 12.5      | WASP-67 b  | 2011   |
| WASP-122  | 07 13 12.35 | -42 24 35.11 | 251.93  | G4     | 11.0      | WASP-122 b | 2015   |
| WASP-140  | 04 01 32.54 | -20 27 03.91 | 180     | K0     | 11.1      | WASP-140 b | 2016   |

## Observation

In order to acquire the initial parameters from the Exoplanet Transit Database (ETD), we used the website's raw data, which is obtained from the observations of CoRoT-12 b, HAT-P-52 b, HAT-P-57 b, HATS-28 b, HATS-34 b, KELT-3 b, WASP-61 b, WASP-67 b, WASP-122 b, and WASP-140 b exoplanets. All of these observations were done through the four filters of B, V, R and Clear. We were required to adjust the data and categorize them in three columns. The first column contains time information that we got from ETD in either Julian Dates (JD) or Heliocentric Julian Dates (HJD), and then changed to Barycentric Julian Dates (BJD) by using Jason Eastman's website. In this part, we also needed observation sites coordinates and star coordinates, which we obtained from ETD and SIMBAD. In the second column, we transformed DMag into Flux by PHOEBE (legacy) 0.32.  Therefore, we were able to attain the normalized data.

In the Table 2, you can perceive information about the tools used in observations. It includes their type and size in centimeters. The average size of tools is 43.54. CCD was used in all of the observations. The Figure 1 shows a sample of observed and theoretical light curve.

**Table 2. Observational specifications and their tools.**

| Planet Name | Observer | Observation's Site | Optic size (cm) | CCD Model | Filter |
|-------------|----------|--------------------|-----------------|-----------|--------|
| CoRoT-12 b  | F. Lomoz | 15°E, 50°N | 25.4 | G2-8300 | Clear |
| CoRoT-12 b  | F. Grau Horta | 1°E, 42°N | 30.5 | CCD FLI PL1001E-1 | R |
| HAT-P-52 b  | Y. Jongen | 5°E, 44° | 43.1 | SBIG STLX11002 | Clear |
| HAT-P-52 b  | F. Campos | 2°E, 41°N | 35.5 | SBIG ST-8XME | Clear |
| HAT-P-52 b  | R. Naves | 2°E, 41°N | 30.5 | CCD Moravian G4-9000 | Clear |

---

[6] http://brucegary.net/AXA/x.htm
[7] http://nsted.ipac.caltech.edu
[8] http://var2.astro.cz/EN/tresca

| | | | | | |
|---|---|---|---|---|---|
| HAT-P-57 b | F. Lomoz | 15°E, 50°N | 30 | ST2000XM | V |
| HAT-P-57 b | A. Wünsche | 5°E, 44°N | 82 | CCD FLI PL230 | V |
| HATS-28 b | P. Evans | 200°E, -21°N | 25 | ST9XE | Clear |
| HATS-34 b | Y. Jongen | 289°E, -30°N | 43.1 | Moravian 4G | Clear |
| KELT-3 b | S. Gudmundsson | 345°E, 64°N | 30 | SBIG STL 11k | Clear |
| KELT-3 b | A. Ayiomamitis | 23°E, 38°N | 30.5 | SBIG ST-10XME | Clear |
| WASP-61 b | C. Quiñones, et al. | 296°E, -31°N | 154 | CCD | B |
| WASP-67 b | P. Evans | 200°E, -21°N | 25 | ST9XE | Clear |
| WASP-122 b | Y. Jongen | 289°E, -30°N | 43.1 | Moravian 4G | V |
| WASP-140 b | F. Lomoz | 15°E, 50°N | 25.4 | G2-8300 | Clear |

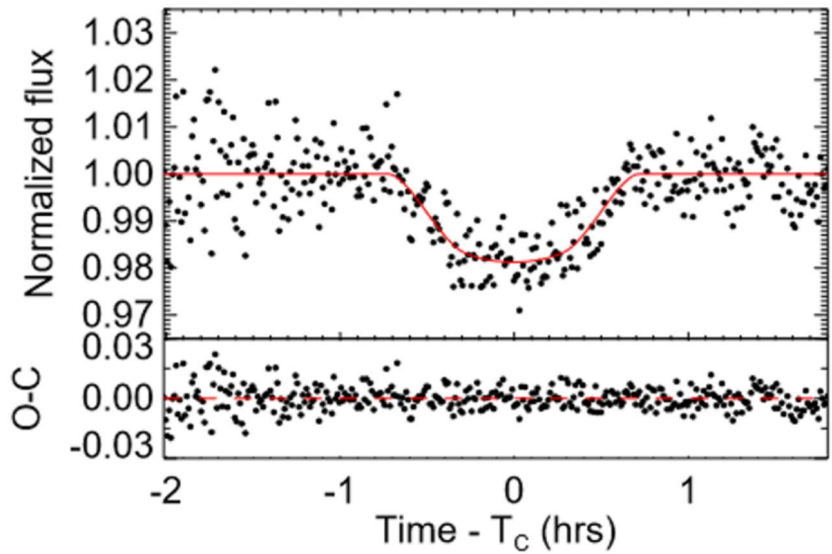

**Figure 1. An example of an observational and theoretical light curve provided by Exofast software for WASP-140 b (Clear filter).**

## Method

In this study, we used Exofast (Fast Exoplanetary Fitting Suite in IDL)[9] software to obtain planetary parameters. This software is an Online Exoplanet observed data analyzer that works based on NASA's exoplanet archive. Jason Eastman programmed this software in 2013, with the support of the California Institute of technology, in partnership with his information team, as an online analyzer robot in IDL (Eastman, 2013).

This software is one of the NASA Exoplanet Archive achievements, which functions by input data. NASA's Exoplanet archive is a great collection of tables that contains not only numeral data of Exoplanets but also Kepler exoplanets candidates such as stellar parameters including positions, magnitudes, temperatures, and etc. Moreover, exoplanet parameters such as mass, orbital parameters, and also discovery and data characterization demonstrated as published radial velocity curves, photometric light curves, images, and spectra (Akeson, 2013, and Eastman, 2019).

Collecting this archive has been started since 1995 by starting more discoveries and information of exoplanets by Mayor and Queloz the cooperating of them group and supporting NExScI (NASA exoplanet science institute) is still in progress. This information is provided by the Kepler-the space telescope that was sent an orbit of earths to discover Exoplanets, in 2009, by NASA, and CoRot (Convection Rotation and planetary transit- Space mission by the French national space study center) and also the given information of research observatories on the earth (Akeson, 2013).

---

[9]https://exoplanetarchive.ipac.caltech.edu/cgi-bin/ExoFAST/nph-exofast

Exofast software receives the name of the Exoplanet and observational data in format of BJD_TDB and Flux. BJD_TDB, which calculated from the online website[10], and normalized flux is obtained by using Phoebe 0.32 software.

It provides the result of analysis as an output that contains information such as the orbital period of planets, inclination, radius, and other parameters of the transit.

## Analysis and conclusion
### A) Calculating the mass

An exoplanet possesses many important properties, one of them being its mass. Knowing the mass of an exoplanet can help us determine other properties of the planet. As an instance, it can lead to a better understanding of the planet's composition by using the mass-radius relations. The mass of an exoplanet is mostly calculated by the Doppler shift (Swift, 2010). The Doppler (or radial velocity) technique, is a method that measures the changes in the motion of a star caused by the gravitational tug of an orbiting planet. Visible-light spectrum of the star demonstrates these changes. Another way of calculating this property is by analyzing the transition of the planet. As the planet transits its star, the alteration in the brightness of the star, determine the ratio between the planet radius to the star radius. Therefore, this information can be placed in a suitable formula to calculate the mass of the exoplanet (Weiss, 2016).

In addition to the methods mentioned above, also another technique can be used. This technique is using the data produced by transits and Kepler's third law. Kepler's third law of planetary motion (published by Johannes Kepler between 1609 and 1619) states a relation between the orbital period of the planet and the semi-major axis of its orbit (Kepler's Three Laws of Planetary Motion[11]). Since the numbers required by the formula are obtained through the analysis of the transitions (available in NEA), we chose this formula as our method of calculating the planet's mass and acquired all the numbers that were needed from NASA's exoplanet archive. The first step to calculate the mass is to determine the host star's mass. Next, the orbital period and the semi-major axis of the planet's orbit should be established. Finally, the mass can be calculated using the following formula:

$$\frac{P^2}{4\pi^2} = \frac{a^3}{G(M+m)} \qquad (1)$$

Where $P$ (s) is the orbital period, $a$ is the Semi-Major Axis (m), $M$ (kg) is the host star's mass, and $m$ (kg) is the planet's mass.

Figure 2 shows the relationship between SMA (AU) and the orbital period of the planets in this study.

---

[10]http://astroutils.astronomy.ohio-state.edu/time/
[11] https://www-spof.gsfc.nasa.gov/stargaze/Kep3laws.htm

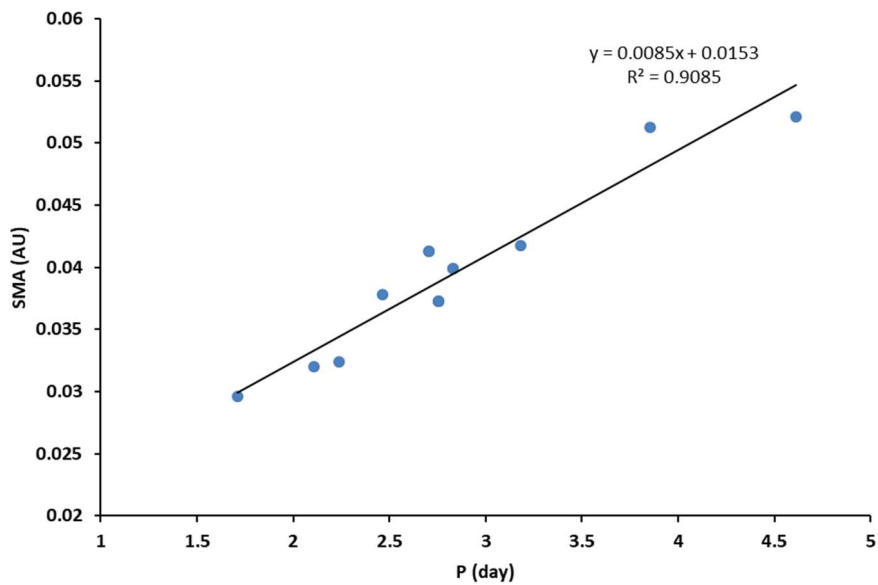

**Figure 2. The relationship between SMA (AU) and the orbital period of the planets, based on the observational parameters obtained.**

To calculate the total mass of stars and planets, Visual Basic language software was designed (Figure 3). Features of this software: Converting required units, calculating the mass of stars and planets, and comparing the calculated mass through observation with the mass calculated based on reference. The results are shown in Table 3.

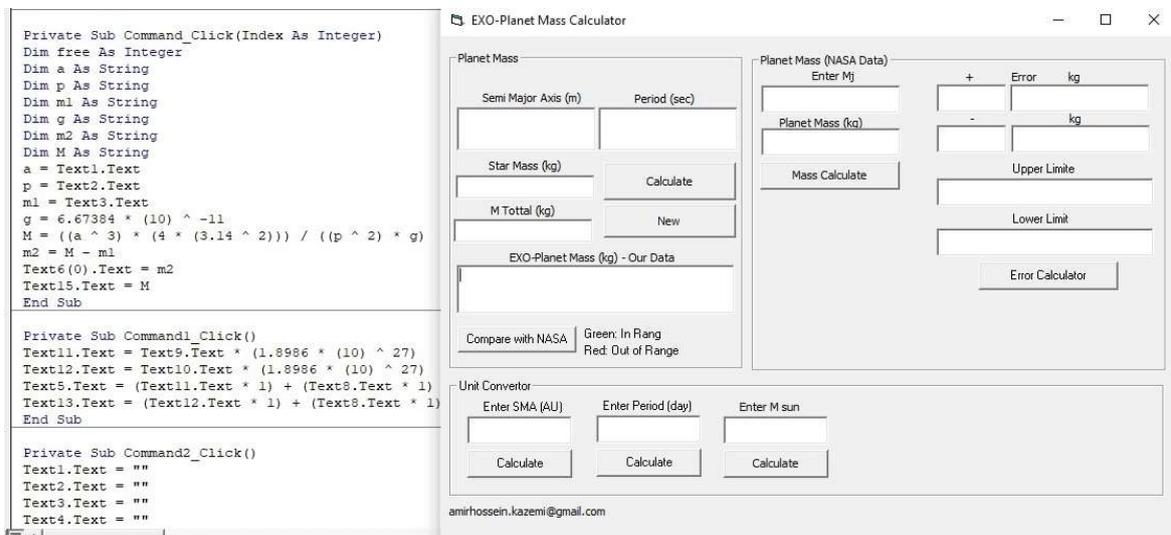

**Figure 3. An image of the software environment designed to calculate the total mass.**

**Table 3. Calculations of the total mass by the software designed in this study.**

| Planet | Semi-Major Axis (m) | Period (sec) | Total Mass (kg) |
|---|---|---|---|
| CoRoT-12 b | 5971647802.6026 | 239252.3532 | 2.19843494365844E+30 |
| CoRoT-12 b | 5971647802.6026 | 239252.3532 | 2.19843494365844E+30 |
| HAT-P-52 b | 5570426313.3852 | 232954.137 | 1.88220942193734E+30 |
| HAT-P-52 b | 5570127117.6438 | 232954.137 | 1.88190614956745E+30 |
| HAT-P-52 b | 5570127117.6438 | 232954.137 | 1.88190614956745E+30 |
| HAT-P-57 b | 899472755762.303 | 208564.38 | 9.88615146224498E+36 |
| HATS-28 b | 6242719144.311 | 269119.1988 | 1.98506862337482E+30 |
| HATS-34 b | 4783990307.1153 | 178181.2206 | 2.03793585713948E+30 |
| KELT-3 b | 6171959351.4699 | 228706.794 | 2.65616185953124E+30 |
| KELT-3 b | 6171809753.5992 | 228706.794 | 2.65596872158041E+30 |
| WASP-61 b | 7671528407.3667 | 326209.14 | 2.5072459528813E+30 |
| WASP-67 b | 7798536999.591 | 390379.932 | 1.83911099012456E+30 |
| WASP-122 b | 4432285713.0996 | 144670.8222 | 2.45846959546179E+30 |
| WASP-140 b | 4845325434.1023 | 189164.1618 | 1.87860135329608E+30 |
| HAT-P-57 b | 5652256348.6581 | 208564.38 | 2.45317635666998E+30 |

**B) The correlation between the stellar metallicity and planet mass**

One of the parameters of stars, named Metallicity is used for measuring the abundance of elements heavier than hydrogen and helium (Ghezzi, 2010). Many researchers, using various methods, have studied stellar metallicity. The stellar metallicity correlates with its planet, which is very significant for exoplanets' investigations. For instance, this important correlation helps us to comprehend the planets' formation and evolution. In addition, studying stellar metallicity can help us to predict the existence of the exoplanets (Wang, 2009).

During the investigations, scientists found another correlation between the stellar metallicity and its planets' mass. The correlation between the stellar metallicity and planet mass is different for planets under different conditions (short or long period planets, massive or low mass planets). The investigations' results show that the rate of occurrence for gas-giant planets, terrestrial planets and gas dwarf planets, is more around the stars with high metallicity (Wang, 2009). The Sub-Jupiter mass planets are predominantly around the stars, which are less metallic than the Jupiter-Mass planets' hosting stars. For instance, the metal-rich stars may have massive Jupiter-Mass planets more (Adibekyan, 2019). On the other hand, the result shows that we have a relationship that is between the stellar metallicity, planet period, and planet mass. Furthermore, one of the researchers demonstrates the Metallicity-Period-Mass diagram for low mass planets and adapted formula for their own diagram in order to show the relation between the stellar metallicity, planet period and planet mass (Sousa, 2018). Moreover, the rate of occurrence for giant planets which are with periods shorter than 10 years and with masses which are fifty times higher than the earth's mass has been estimated to be 13.9±1.7% (Mayor, 2011). According to the investigations' results, a positive correlation exists between the planet mass and the planets' hosting star metallicity.

**C) H-R diagram for the host stars**

The positions of all host stars in this study, in which the theoretical Zero Age Main Sequence (ZAMS) and Terminal Age Main Sequence (TAMS) are shown in the H-R diagram in Figure 4 (The results are shown in Table 4).

**Table 4. Calculations of host stars and parameters determining their position in the H-R diagram.**

| Planet Name | T (K) | Log T | L | Log L |
|---|---|---|---|---|
| CoRoT-12 b | 5675 | 3.753 | 1.300 | 0.114 |
| HAT-P-52 b | 5131 | 3.710 | 0.657 | -0.182 |
| HAT-P-57 b | 7500 | 3.875 | 3.851 | 0.585 |
| HATS-28 b | 5498 | 3.740 | 0.772 | -0.112 |
| HATS-34 b | 5380 | 3.730 | 0.851 | -0.070 |
| KELT-3 b | 6304 | 3.799 | 2.385 | 0.377 |
| WASP-61 b | 6250 | 3.795 | 2.005 | 0.302 |
| WASP-67 b | 5200 | 3.716 | 0.614 | -0.211 |
| WASP-122 b | 5730 | 3.758 | 3.246 | 0.511 |
| WASP-140 b | 5260 | 3.720 | 0.691 | -0.160 |

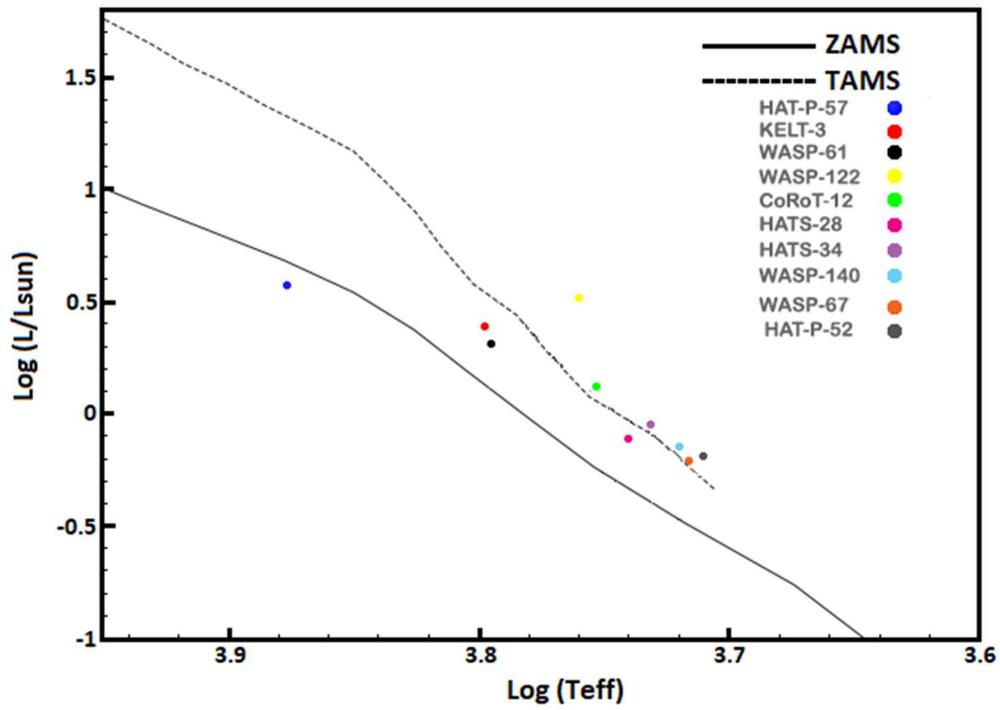

**Figure 4. The position of the host stars on the H-R diagram.**

## Acknowledgments

IOTA/ME held this project as a scientific activity on exoplanets for students in the spring of 2020.

## Appendix

Observational and theoretical light curves are included in this study, presented in the appendix.

### CoRoT-12 b (R)

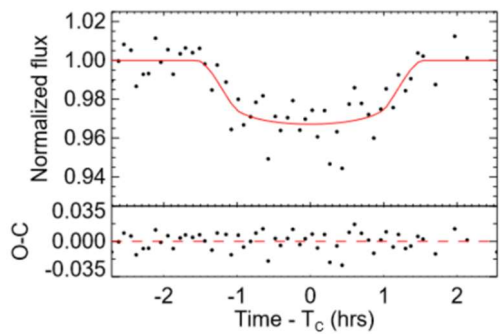

### HAT-P-52 b (Clear)

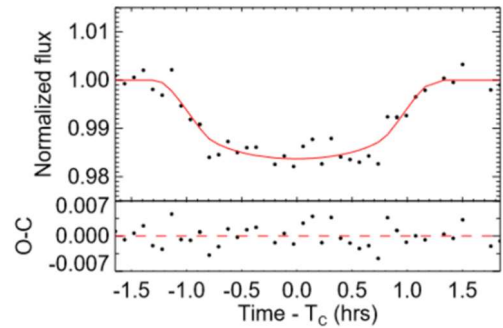

### HAT-P-52 b (Clear)

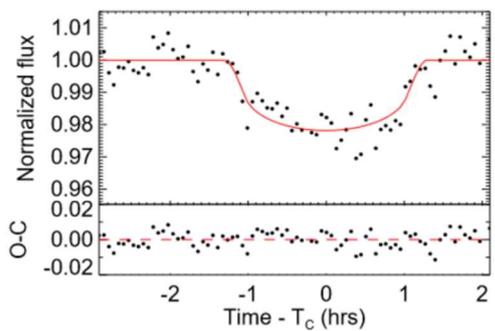

### HAT-P-52 b (Clear)

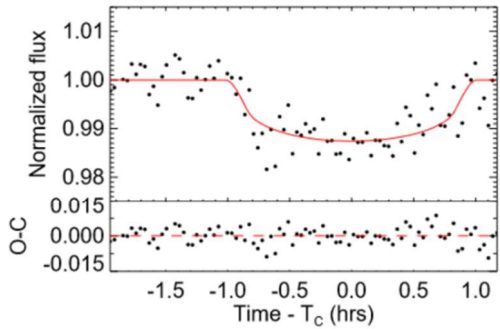

### HAT-P-57 b (V)

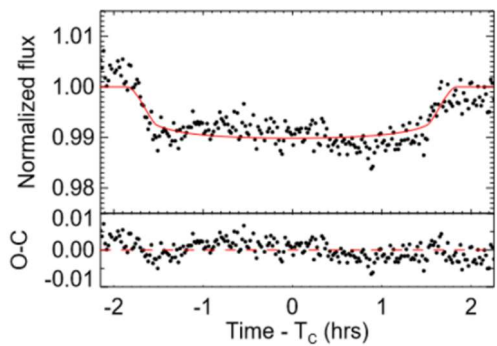

### HAT-P-57 b (V)

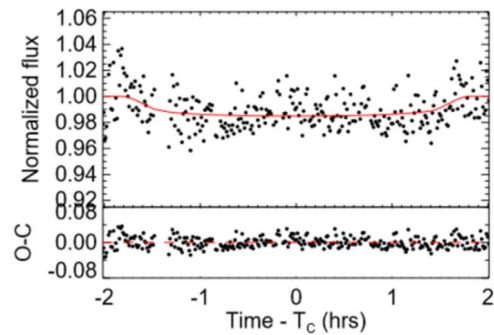

### HATS-28 b (Clear)

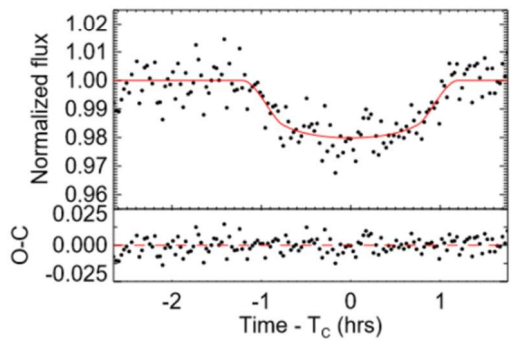

### HATS-34 b (Clear)

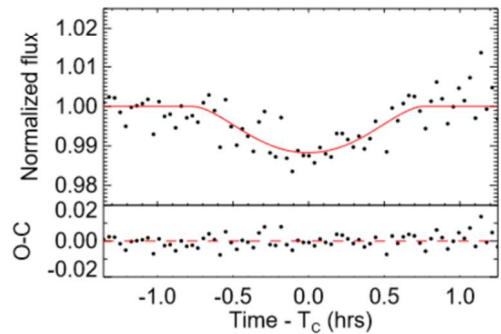

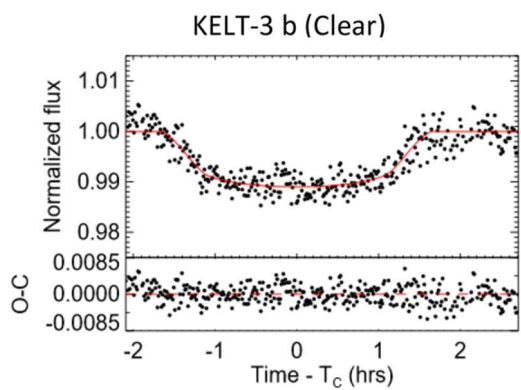

KELT-3 b (Clear)

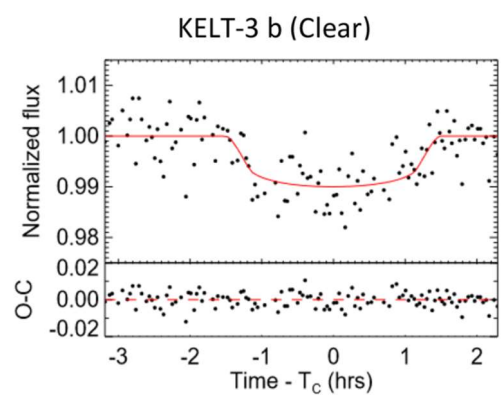

KELT-3 b (Clear)

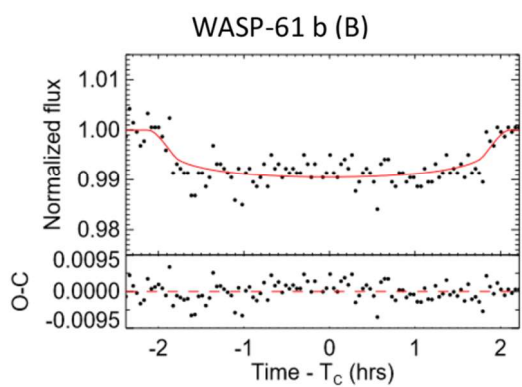

WASP-61 b (B)

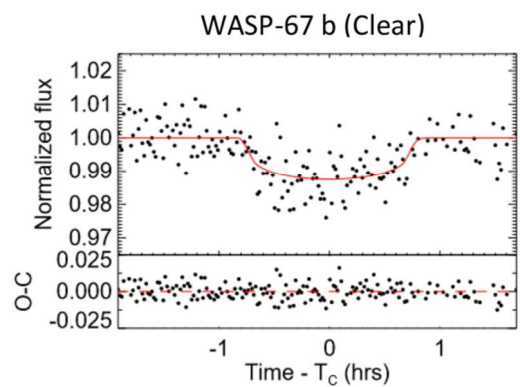

WASP-67 b (Clear)

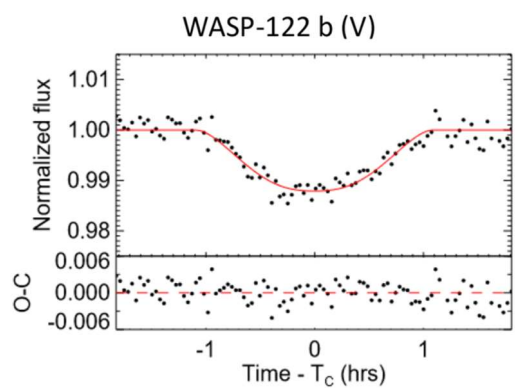

WASP-122 b (V)